\documentclass[trackchanges,twocolumn]
{aastex7}

\usepackage{booktabs}
\usepackage{array}
\usepackage{float}
\usepackage{amsmath}
\usepackage{graphicx}
\usepackage{placeins}
\usepackage{subcaption}
\begin{document}

%\title{\textbf{IGR J12580+0134: A Candidate Repeated Partial Tidal Disruption Event Inferred from Late-Time Radio Rebrightening}}

\title{\textbf{IGR J12580+0134: A Possible Repeated Partial Tidal Disruption Event Inferred from Late-Time Radio Re-brightening}}

\author[0009-0000-0635-5679]{Po Ma}
\affiliation{Department of Astronomy, School of Physics, Huazhong University of Science and Technology, Luoyu Road 1037, Wuhan, 430074, China}
\email[]{}

\author[0009-0002-7730-3985]{Shao-Yu Fu}
\affiliation{Department of Astronomy, School of Physics, Huazhong University of Science and Technology, Luoyu Road 1037, Wuhan, 430074, China}
\email[show]{syfu@hust.edu.cn}

\author[0000-0003-3454-6522]{Linhui Wu}
\email[show]{wulinhui2@gmail.com}
\affiliation{Shanghai Astronomical Observatory, Chinese Academy of Sciences, 80 Nandan Road, Shanghai, 200030, P. R. China}

% \author{Hong-Zhou Wu}
% \affiliation{Department of Astronomy, School of Physics, Huazhong University of Science and Technology, Luoyu Road 1037, Wuhan, 430074, China}

\author[0000-0003-3440-1526]{Wei-Hua Lei}
\affiliation{Department of Astronomy, School of Physics, Huazhong University of Science and Technology, Luoyu Road 1037, Wuhan, 430074, China}
\email[]{leiwh@hust.edu.cn}

\author[0000-0003-4891-3186]{Qiang Yuan}
\affiliation{Key Laboratory of Dark Matter and Space Astronomy, Purple Mountain Observatory, Chinese Academy of Sciences, Nanjing 210023, China}
\affiliation{School of Astronomy and Space Science, University of Science and Technology of China, Hefei 230026, China}
\email[]{}

\begin{abstract}
Repeating partial tidal disruption events (pTDEs) provide a direct probe of stellar orbits and episodic mass loss around supermassive black holes, but robust identification requires multi-band and multi-epoch evidence.
%consistent with a single physical origin. 
We investigate whether the late-time radio rebrightening of the nuclear transient IGR~J12580+0134 in NGC~4845 can be explained as a repeating pTDE, using multi-epoch Karl G.\ Jansky VLA observations together with X-ray constraints from \textit{Swift}/XRT and \textit{NICER}. Through a systematic analysis of the radio data, we identify two well-defined radio flares and a possible third late-time rebrightening flare. Modeling the second flare with a synchrotron afterglow framework using Markov Chain Monte Carlo fitting is consistent with a sub-relativistic outflow with a characteristic velocity of order ${v \simeq 0.3c}$, an isotropic-equivalent kinetic energy of order ${10^{50}}$ erg, and an approximately constant-density circumnuclear medium. No significant contemporaneous brightening is detected by \textit{Swift}/XRT during the 2016 radio flare, while faint \textit{NICER} flares in 2023 suggest intermittent low-level accretion. We also considered several possible interpretations for the late-time radio rebrightening, and found that the repeated pTDE scenario provides a more natural overall explanation for the observed phenomenology. Given the currently sparse data coverage, continued sensitive radio and X-ray monitoring will be essential to test this interpretation and to search for future reactivations.
\end{abstract}
%\uat{Black holes}{162};\uat{Supermassive black holes}{1663}; \uat{Radio active galactic nuclei}{2134}

%, estimated from the earliest observed epoch of the first radio flare and the fiducial onset time adopted for the second radio flare

\section{INTRODUCTION}\label{sec:intro}
When a star approaches the tidal radius of a supermassive black hole (SMBH), it can be fully disrupted by tidal forces. A fraction of the stellar debris is subsequently bound and falls back, producing a luminous flare whose bolometric luminosity and decline rate encode the nature of the disrupted body and the efficiency of circularisation \citep{Rees1988}. Such events, called tidal disruption events (TDEs), provide a unique method to study accretion and jet-launching physics around SMBHs of quiescent galaxies \citep{Wang2025SciA}. 

A tidal encounter does not necessarily lead to a full disruption. If the stellar pericentre is sufficiently large, the star can survive the passage and lose only a fraction of its mass, resulting in a similar flare. In this case, a partial tidal disruption event (pTDE) happens \citep{Wevers2023ApJ...942L..33W,Somalwar2025-2020vdq,Guillochon_2013,Coughlin_2019}. The encounter strength is commonly quantified by the penetration factor, $\beta \equiv R_{\rm t}/R_{\rm p}$, where $R_{\rm p}$ is the pericentre distance. In recent years, numerical simulations have shown that the critical $\beta$ values separating the onset of partial disruption from full disruption depend sensitively on the internal structure of the star \citep[e.g.,][]{Guillochon_2013, Ryu_2020}. Several theoretical studies suggest that partial disruptions may occur at rates similar to, or exceeding, those of full disruptions, implying that pTDEs could make an important contribution to the total TDE population \citep[e.g.,][]{stone_metzger2016, Stone:2020Rates, chen-shen2021,zhong2022}. Moreover, numerical simulations predict that pTDEs can exhibit fallback-rate evolutions distinct from those of full disruptions, potentially imprinting observable differences on the light-curve decline. For full disruptions, the fallback rate is expected to approach the canonical late-time decay, $\dot{M}_{\rm fb}\propto t^{-5/3}$ \citep{Rees1988, phinney1989}, in contrast, numerical simulations suggest that partial disruptions can exhibit steeper fallback decays, approaching $\dot{M}_{\rm fb}\propto t^{-9/4}$ in certain regimes \citep{Guillochon_2013, Coughlin_2019, Ryu_2020}.

While the fallback-rate evolution may provide a useful diagnostic, it can be challenging to identify pTDEs from a single flare alone. A more distinctive signature is offered by systems that produce multiple episodes of nuclear flaring. Such repeated pTDEs can arise when a star is placed on a bound, highly eccentric orbit around an SMBH and undergoes partial stripping at successive pericentre passages \citep{Somalwar2025-2020vdq,payne2021asassn,Liu2025}. One possible formation channel is the Hills mechanism, in which a stellar binary is tidally separated during a close passage by the SMBH, ejecting one component as a hypervelocity star while capturing the other onto a tight orbit \citep{hills1988hyper}. The captured star may then experience repeated partial disruptions as it returns to pericentre , producing a sequence of flares separated by months to years\citep{cufari2022using, Lu2023}. Such repeated events offer a direct observational pathway to establishing partial disruptions, since recurrence naturally follows from the survival of the star after the initial passage.

%The surviving stellar core can then undergo repeated partial stripping at subsequent pericentre passages, producing a sequence of flares separated by months to years. Such repeated events offer a direct observational pathway to establishing partial disruptions, since recurrence naturally follows from the survival of the star after the initial passage.
Observationally, repeated pTDE candidates are typically characterised by multi-episode nuclear flaring activity with recurring timescales, often accompanied by similarities in photometric and spectroscopic properties between episodes. Nevertheless, confirming repeated pTDEs can be non-trivial, since alternative scenarios—such as double TDEs from stellar binaries, SMBH binaries, or enhanced nuclear transient rates in dense stellar environments—may also produce multi-episode flaring behaviour \citep[e.g.,][]{Mandel_2015, wu-yuan2018, Arcavi2014, Hammerstein2023ApJ, wang2024, Pfister2020}.

To date, only a small number of repeated pTDE candidates have been reported in the literature. These sources are characterised by multi-episode nuclear flares recurring on timescales from days to years, and some exhibit similar photometric and/or spectroscopic properties across different episodes. We summarize their basic observational properties in Table~\ref{tab:rptde}. 
Importantly, the best-studied repeated pTDE candidates have generally been identified through recurrent optical/UV and/or X-ray nuclear flaring, often with similarities in their photometric or spectroscopic properties across different episodes. In particular, systems such as AT2022dbl and AT2020vdq provide useful examples for comparison, because their repeated activity has been characterized in considerably greater multi-wavelength detail. For AT2022dbl, recurrence is inferred mainly from the optical/UV behavior: the two flares are separated by about two years and exhibit remarkably similar rise times, temperatures, peak luminosities, and broad optical spectral features. Neither flare shows detectable X-ray emission, although weak X-ray emission is observed between the two events. \citep{hinkle2025doubleluminousflaresnearby,makrygianni2025doubletidaldisruptionevent}. By contrast, AT2020vdq was originally identified as an optical- and radio-flaring TDE, but its later repeated activity was marked by a dramatic optical rebrightening about 947 days after discovery, accompanied by extremely broad optical/UV spectral features and faint X-ray emission, with no evidence for a newly launched radio-emitting outflow \citep{Somalwar2025-2020vdq}. These sources show that repeated pTDE candidates do not share a comment multi-wavelength template, and that the strongest observational evidence for recurrence has so far often come from repeated optical/UV flaring and spectroscopic similarity rather than from radio behaviour alone.
% Taken together, these sources show that repeated pTDE candidates can display diverse multi-wavelength signatures, and that the clearest evidence for recurrence often comes from repeated optical/UV flaring and spectroscopic similarity rather than from radio behaviour alone.

Radio observations reveal a diverse phenomenology in TDEs. The radio emission is generally interpreted as synchrotron radiation produced as outflowing material interacts with the circumnuclear medium (CNM), with inferred outflows ranging from luminous relativistic or mildly relativistic jets in a small subset of events to slower non-relativistic outflows in the majority of radio-detected non-jetted TDEs \citep{alexander2020radio}. Although some TDEs are detected in the radio soon after discovery, more recent late-time monitoring has shown that delayed radio brightening and rebrightening on timescales of several hundred to several thousand days are not uncommon. In particular, in a systematic study of optically selected TDEs, \citet{cendes2024ubiquitouslateradioemission} found that roughly 40\% of events are detected in the radio at late times, inferred typical outflow velocities of $\sim 0.02$--$0.15c$, and argued that delayed outflows are generally more plausible than a population dominated by off-axis relativistic jets. More recently, \citet{zhou2026investigatingcircumnuclearmediumtidal} compiled one of the most comprehensive samples of radio-observed TDEs to date and showed that the temporal and spectral evolution of radio TDEs can be used to constrain the density profile of the CNM, further highlighting radio observations as a powerful probe of both TDE outflows and their environments. Individual events further show that late-time radio activity can take a variety of forms, including delayed radio flares in ASASSN-15oi \citep{Horesh_2021}, a late radio/X-ray flare in AT2019azh \citep{Sfaradi_2022}, a second radio flare in AT2020vwl \citep{goodwin2024secondradioflaretidal}, and rapidly evolving double-peaked radio behavior in AT2024tvd \citep{sfaradi2025radiobrightoffnucleartde2024tvd}. These examples make clear that late-time radio brightening is not uncommon in TDEs. Delayed emission associated with candidate SMBHB systems has also been discussed at other wavelengths. \citet{shu2020xrayflares} reported delayed X-ray brightening in the TDE OGLE16aaa about 140 days after the optical outburst, which they interpreted as possibly arising from a supermassive black hole binary or patchy obscuration. In the radio band, a related delayed flare has been reported in the SMBHB candidate SDSS J143016.05+230344.4, where the emission was attributed to kinetic-energy dissipation in a compact outflow or jet-base disturbance propagating through a structured CNM \citep{an2026delayedradioflaretraces}.

IGR J12580+0134 is a TDE that occurred in 2011 in the nucleus of the nearby galaxy NGC 4845 with a redshift distance of 18 Mpc \citep{walter2011igrj12580+,Lei_2016,yuan2016catching,Yu2022}. Discovered by \emph{INTEGRAL}, the event exhibited a sub-Eddington X-ray flare consistent with the tidal disruption of a super-Jupiter--mass object \citep{walter2011igrj12580+,nikolajuk2013tidal}. A compact radio counterpart was detected later that year, showing rapid brightening and subsequent evolution consistent with synchrotron emission from an expanding outflow \citep{Irwin2015,perlman2017}.

A late-time radio-only flare was reported more than a decade after the original event \citep{Perlman2022}. This rebrightening was interpreted as resulting from the interaction between the initial outflow and dense circumnuclear material \citep{perlman2017,Perlman2022}. Here, we argue that the recurrence timescale and energetics are instead more naturally explained by a repeated partial tidal disruption event (pTDE), identifying IGR J12580+0134 as a candidate for a recurring TDE system.

This paper is organized as follows. Section 2 describes the observational data and data reduction procedures. In Section 3, we present the theoretical framework for the repeating tidal disruption scenario. Section 4 provides the main results, where the physical parameters of the outflow are constrained from radio observations using Markov Chain Monte Carlo (MCMC) methods. Finally, Section 5 summarizes our findings and discusses their implications for interpreting IGR J12580+0134 as a candidate repeating partial tidal disruption event.

\section{Observations and Data Reduction}\label{sec:Observations}
\subsection{Radio Data}

\setcounter{footnote}{0}
In this work, we analyze radio observations of IGR J12580+0134 obtained with the Karl G. Jansky Very Large Array (VLA) spanning the period from 1995 to 2024, covering frequencies in the L, S, C, and X bands. The flux density of IGR J12580+0134 is derived using three approaches: (1) directly adopting measurements from \citet{Irwin2015} and \citet{Perlman2022}; (2) measuring flux densities from the \textit{Very Large Array Sky Survey}\footnote{https://vlass-dl.nrao.edu/vlass/quicklook/} (VLASS) images; and (3) analyzing two additional datasets retrieved from the NRAO archive\footnote{https://data.nrao.edu/portal/} (project codes: 21A-033 and 22B-016). In the following, we briefly summarize the VLA data reduction procedures applied to these two additional datasets.

We retrieved these two calibrated datasets from the NRAO archive, which had been processed using the automated VLA calibration pipeline in CASA (v6.6.1; \citet{McMullin2007}). In the pipeline, 3C286 was used for flux density and bandpass calibration, while J1224+0330 served as the calibrator for time-dependent complex gains. We then performed imaging using the TCLEAN task in CASA (v6.5.4). We employed the multi-scale multi-frequency synthesis (MS-MFS) mode \citep{Rau2011} to better recover extended and faint emission, and adopted a Briggs weighting scheme with a robust parameter of 0 (or 0.5) to balance resolution and sensitivity.

Following \citet{Irwin2015}, we adopt a host-galaxy disk component of
${7 \pm 1}$ mJy in the C-band and ${19 \pm 4}$ mJy in the L-band, together
with a spectral index of ${-0.74}$ for this component. Using this spectral
index, the corresponding galaxy disk contribution in the S-band is interpolated
as ${\sim 11.8}$ mJy. For the S-band data presented in this work, we
subtract this estimated host-galaxy disk contribution to isolate
the flux density of the AGN component. Radio observations are presented in Table~\ref{tab:obserinf}, with the data obtained and reduced in this work highlighted in red. Based on these newly analyzed radio data, we identify a possible third late-time radio rebrightening episode.

\subsection{X-ray Data}

IGR J12580+0134 was first identified as a luminous X-ray flare in the nucleus of NGC 4845, consistent with a TDE. The event was traced by follow-up \textit{XMM-Newton} observations in January 2011 and is widely interpreted as the disruption of a substellar object by a low-mass supermassive black hole \citep{nikolajuk2013tidal}.  

Spectral analyses of the 2011 observations reveal a heavily absorbed power-law continuum with photon index $\Gamma \approx 2.2$ and hydrogen column density $N_{\rm H} \sim 7\times10^{22}\,\mathrm{cm^{-2}}$, characteristic of an accreting source.  
The average 2--10 keV flux reached ${\sim6\times10^{-11}\,\mathrm{erg\,cm^{-2}\,s^{-1}}}$ during the flare \citep{nikolajuk2013tidal}.   
Moreover, the X-ray light curve follows the canonical $t^{-5/3}$ decline expected from fallback accretion after tidal disruption, providing strong support for the TDE interpretation \citep{Rees1988}.

\begin{figure*}[htbp]
  \centering 
  \includegraphics[width=0.98\textwidth]{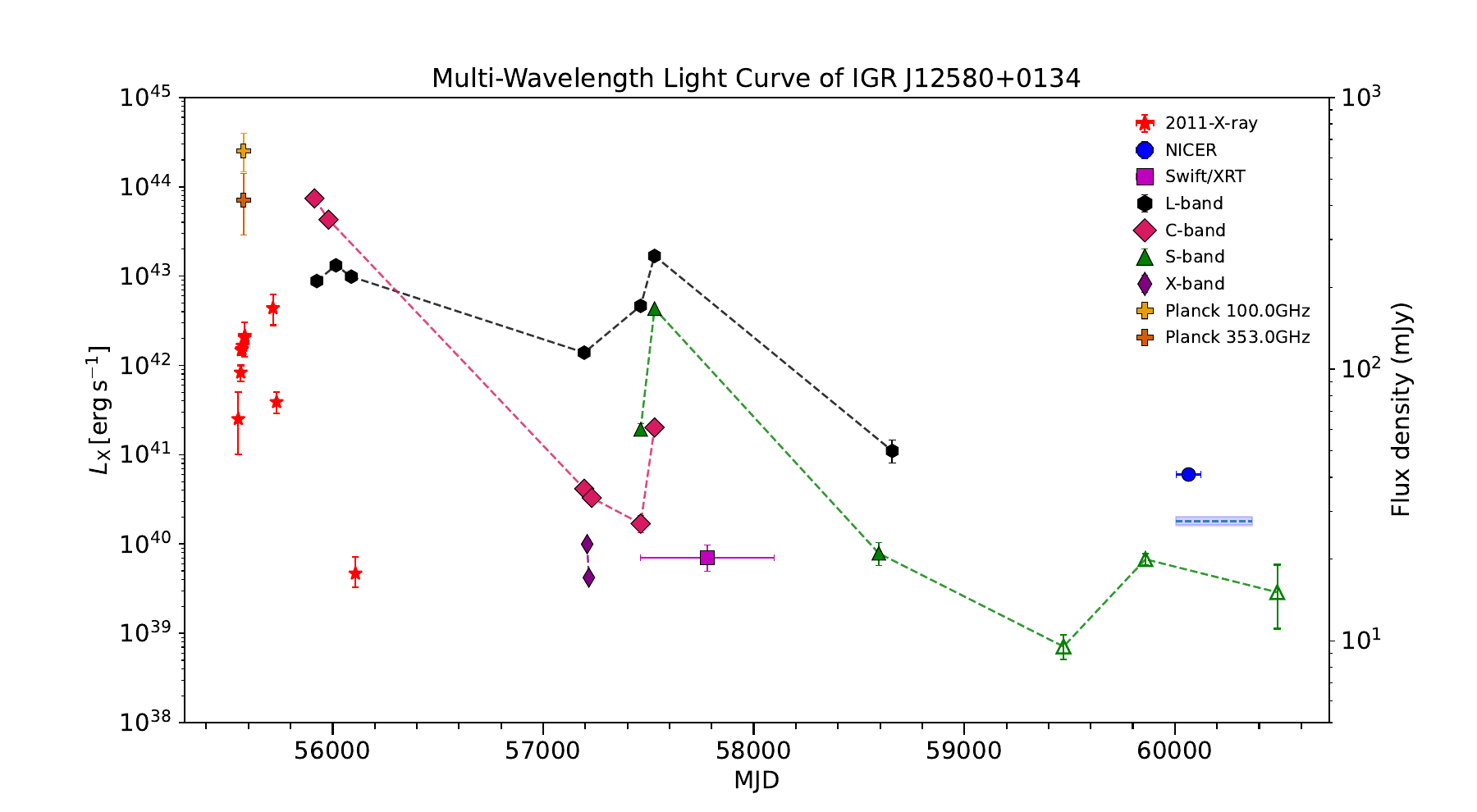}
  \caption{Multi-wavelength light curves of IGR~J12580+0134, combining X-ray and radio data. Radio flux densities are plotted for  L-band (black hexagons), C-band (pink diamonds), S-band (green triangles), and X-band (purple thin diamonds). Red stars represent the 2011 X-ray flare in the 17.3--80 keV band (data from \citet{nikolajuk2013tidal}). The Swift/XRT observation (0.5--7 keV) is shown as magenta squares. A faint X-ray activity detected in 2023 via NICER (0.4--10 keV) monitoring is indicated by blue circles, and the shaded region indicating the background flux level between 2021 June 15 and 2024 February 28 (data from \citet{Danehkar_2025}). Early submillimeter measurements from \textit{Planck} at 100 and 353 GHz are shown as golden and dark-orange crosses (data from \citet{yuan2016catching}).
  The open triangle symbol denotes the radio data obtained from our analysis, including flux densities measured from VLASS images and from our uniform re-analysis of the archival VLA datasets 21A-033 and 22B-016.} The fluxes are plotted on a logarithmic scale to highlight the dynamic range of the event.
\label{fig:multi-wave}
\end{figure*}

Despite continued multiwavelength monitoring, \citet{Perlman2022} analyzed new late-time \textit{Swift}/XRT observations obtained during the 2016 radio flare, together with archival \textit{Swift} and \textit{XMM-Newton} data from \citet{nikolajuk2013tidal}. They reported no statistically significant X-ray brightening during the 2016 radio flare. Because the individual observations had limited counts, \citet{Perlman2022} co-added the final three Swift datasets to improve the signal-to-noise ratio, producing a single flux measurement that we adopt and display in Figure~\ref{fig:multi-wave}.

More recently, Neutron star Interior Composition
Explorer (hereafter \textit{NICER}) monitoring has revealed faint X-ray flares during March–June 2023 that are substantially weaker than the 2011 outburst. These low-luminosity events are likely associated with extremely weak accretion onto the central black hole and may indicate the presence of a low-luminosity active galactic nucleus \citep{Danehkar_2025}. Interestingly, the timing of these late-time X-ray flares 
broadly overlaps with the possible third radio rebrightening episode identified in 
our radio analysis. 
We also adopt these \textit{NICER} measurements and display them in Figure~\ref{fig:multi-wave}.

\section{Modeling} \label{sec:modeling}

% \subsection{The Possibly Orbital Period of IGR J12580+0134}
% For IGR J12580+0134, we estimate the orbital period by adopting the time separation between the radio L-band peaks of the first and second flares. However, strictly speaking, the observed peak separation does not equal the true orbital period. Instead, it obeys
% \begin{equation}
%     \Delta T_{\mathrm{obs}}= T_{\mathrm{orb}} - T_{peak,1} + T_{peak,2},
% \end{equation}
% where $\Delta T_{\mathrm{obs}}$ is the observed separation between the two peaks in the L band (2012-Mar-30 and 2016-May-21),
% $T_{\mathrm{orb}}$ is the orbital period, and $T_{peak,1}$ and $T_{peak,2}$ denote the delays to the
% peak fallback rate following the first and second passages, respectively. The two peak times, $T_{\mathrm{peak},1}$ and $T_{\mathrm{peak},2}$, will generally not be identical, 
% since the surviving stellar core is spun up during each passage
% \citep{Ryu_2020, bandopadhyay2024repeating} and simultaneously loses mass as a result of partial disruption. Here we assume that the peak luminosity traces the peak fallback rate. According to Table~\ref{tab:obserinf}, $\Delta T_{\mathrm{obs}} \approx \Delta T = 1513$ days.
% \setcounter{footnote}{0}
\subsection{A Characteristic Recurrence Timescale}
% \textbf{For IGR~J12580+0134, we use the interval between the onset of the first radio flare and the presumed onset of the second flare to define a characteristic recurrence timescale, rather than a direct measurement of the orbital period. 

In a synchrotron outflow scenario, neither the first detectable radio brightening nor the radio peak necessarily provides a direct measurement of the physical triggering time, since both may be delayed relative to the launch event by synchrotron self-absorption, shock evolution, and the structure of the surrounding medium.

% Specifically, we take MJD~55925 (2011 December 30), corresponding to the epoch of the first radio flare identified by \citet{Irwin2015}, and adopt MJD~57340 as the reference epoch in our modelling, corresponding to the presumed onset of the second radio episode. This gives $\Delta T_{\rm rec,12}\approx 1415~{\rm d}$.
% In addition, the sparsely sampled late-time radio data show a possible third radio rebrightening. If we tentatively take MJD~59471 (2021 September 14) as the onset epoch of this candidate third radio episode, the interval between the presumed onsets of the second and third episodes is
% $\Delta T_{\rm rec,23}\approx 2131~{\rm d}$.
% We treat these intervals as phenomenological recurrence timescales and use them only for order-of-magnitude dynamical estimates, rather than as direct measurements of the true orbital period.

For IGR J12580+0134, we take MJD~55925 (2011 December 30), corresponding to the epoch of the first radio flare identified by \citet{Irwin2015}, as a reference epoch for the first observed radio episode rather than a precisely measured onset time. This is because the rising phase of the first radio flare was not covered by the available observations, and the true onset of the radio activity could have occurred earlier. For the second radio episode, the onset is also not directly observed; it is bracketed by the lower-flux radio measurements in 2015 and the subsequent rebrightening in 2016. We therefore adopt MJD~57340 only as a representative reference epoch in our modelling, with a plausible onset interval of approximately MJD 57231--57529 (from 2015 July 28 to 2016 May 21). Using MJD~55925 as the reference epoch for the first observed radio episode, this gives a phenomenological interval of 
$\Delta T_{\rm rec,12}\simeq 1306$--$1604~{\rm d}$, although the true interval could be longer if the first radio episode started before its first radio detection.
The late-time radio data are even more sparsely sampled, but they show a possible third radio rebrightening. If we associate this candidate episode with the rise between the 2019 April 21 and 2022 October 8 S-band measurements, its onset can be bracketed by MJD~58594--59860. Combining this interval with the plausible onset interval of the second radio episode gives
$\Delta T_{\rm rec,23}\simeq 1065$--$2629~{\rm d}$. The broad overlap between $\Delta T_{\rm rec,12}$ and $\Delta T_{\rm rec,23}$ suggests that the observed radio episodes may be compatible with quasi-recurrent activity. We therefore treat these intervals as phenomenological recurrence timescales, rather than direct measurements of the true orbital period.

For a black hole with mass $M_{\rm BH} \approx 3\times10^{5} - 1.8\times10^{7}\,M_\odot$ \citep{Lei_2016}, 
an orbital period of $\sim${1400} days corresponds to a semi-major axis of
$a \sim ({2.47}\times10^{15}-{9.67}\times10^{15})\,\mathrm{cm}$. Using the stellar velocity dispersion ($\sigma$) of NGC~4845 obtained from the HyperLeda\footnote{\url{http://leda.univ-lyon1.fr/}} database \citep{2014A&A...570A..13M}, 
we estimate the radius of influence of the SMBH as
\begin{equation}
    r_{\rm infl} \simeq \frac{G M_{\rm BH}}{\sigma^2} \approx 
    (2.22\times10^{17} - 1.33\times10^{19})~{\rm cm},
\end{equation}
which is larger than the derived semi-major axis $a$ by roughly two to three orders of magnitude.

Since most disrupted stars originate in this region, tidal dissipation alone is insufficient to bind the star tightly enough to yield an orbital period of {1400} days \citep{cufari2022using}. A plausible dynamical pathway is the Hills mechanism \citep{hills1988hyper}. In this process, the center of mass of a binary star with total mass $M_{\rm b}$ and separation $a_{\rm b}$ approaches the SMBH on an orbit with a pericenter distance smaller than $R_a = a_{\rm b} (M_{\rm BH}/M_{\rm b})^{1/3}$,
at which the binary is tidally disrupted.~One star is ejected from the system, whereas the companion becomes bound to the SMBH with an orbital period of order of years.
%, consistent with the measured value of $\sim$1500 days (see below). 

For mass transfer or partial disruption to occur, the pericenter distance must be comparable to the tidal radius. This corresponds to a penetration factor of $\beta \equiv R_{\rm t}/R_{\rm p} \gtrsim 0.6$ (i.e., $R_{\rm p} \lesssim 2R_{\rm t}$), such that the star begins to lose mass once it passes within approximately twice the tidal radius \citep{10.1093/mnrasl/slac106}.

 Furthermore, the orbital period of the captured star can be estimated from the binding energy imparted during the tidal breakup of the binary. The typical period scales as 
\begin{equation}
    T_\ast \sim \frac{a_b^{3/2}}{\sqrt{G M_b}}
    \left(\frac{M_{\rm BH}}{M_b}\right)^{1/2}.
\end{equation}
%where $M_b$ and $a_b$ are the binary mass and separation, respectively. 
To avoid disruption before reaching $R_t$, the binary separation must satisfy \citep{QUINLAN199635},
\begin{equation}
a_b \lesssim \frac{G M_b}{\sigma^2}.
\end{equation} 
Combining these expressions yields
\begin{equation}
    T_\ast \lesssim \frac{G M_b}{\sigma^3}
    \left(\frac{M_{\rm BH}}{M_b}\right)^{1/2}.
\end{equation}

Using the $M_{\rm BH}-\sigma$ relation for late-type galaxies\citep{2000ApJ...539L..13G, 2000ApJ...539L...9F},
$M_{\rm BH}/M_0=(\sigma/\sigma_0)^n$, with
$M_0=1.17\times10^8\,M_\odot$, $\sigma_0=200~{\rm km~s^{-1}}$,
and $n=5.06$ \citep{McConnell_2013}, the above constraint becomes
\begin{equation}
T_* \lesssim 2077
\left(\frac{M_b}{M_\odot}\right)^{1/2}
\left(\frac{M_{\rm BH}}{M_0}\right)^{-47/506}
~{\rm days}.
\end{equation}

% This suggests that orbital periods of several thousand days are plausible for typical parameters, consistent with the $\sim\textbf{1400}$ days period inferred for IGR J12580+0134.

This suggests that orbital periods of several thousand days are plausible for typical parameters, broadly consistent with the recurrence timescales inferred for IGR~J12580+0134.

Thus, the Hills mechanism provides a natural explanation for capturing the star onto such a tight orbit, making subsequent partial tidal disruptions possible.

% In order to constrain the orbital parameters of the bound star, we follow the analytical framework developed in the context of AT2022dbl \citep{makrygianni2025doubletidaldisruptionevent}. In that work, the authors employed the Hills mechanism together with angular momentum relaxation to derive the eccentricity constraint for stars captured by a SMBH. Building upon this formulation, we apply the same relation to our system and carry out explicit estimates in order to verify the consistency between theoretical expectations and the observed orbital period.

In order to constrain the orbital parameters of the bound star, we adopt the orbital-dynamical framework presented by \citet{makrygianni2025doubletidaldisruptionevent}, who used the Hills mechanism together with angular-momentum relaxation to derive eccentricity constraints for stars captured by a SMBH. Here, we use this framework only to describe the dynamical evolution of a bound stellar orbit. AT2022dbl is used only as a dynamical reference, since it does not show radio re-brightening like that observed in IGR J12580+0134. We apply the same orbital relation to our system solely as a consistency check on whether the inferred recurrence timescale can be accommodated within a bound-orbit scenario.

Specifically, the orbital eccentricity can be expressed as
\begin{align}
&1 - e \notag\\
&\approx \beta_\ast^{-1} (2\pi)^{2/3}
\frac{R_\ast}{\left(G M_\ast T_\ast^2\right)^{1/3}} \notag \\
&\approx {0.0027}
\left(\frac{T_\ast}{{1400}~{\rm d}}\right)^{{-2/3}}
\left(\frac{R_\ast}{R_\odot}\right)
\left(\frac{M_\ast}{M_\odot}\right)^{-1/3}
\left(\frac{\beta_\ast}{0.7}\right)^{-1}
\end{align}
where $\beta_\ast = R_t / R_p$ is the penetration factor, $R_\ast$ and $M_\ast$ are the stellar radius and mass, $T_\ast$ is the orbital period. In our analysis, we adopt $\beta_\ast = 0.7$. 
%, consistent with the result obtained from the MOSFiT fitting below. 
For a solar-type star ($M_\ast = 1 M_\odot$, $R_\ast = 1 R_\odot$) around a black hole of $M_{\rm BH} \sim 10^6 M_\odot$, we obtain an eccentricity $e \approx {0.9973}$ and $R_{\rm p} \approx 10^{13}\ \rm cm$, making the  semi-major axis $a \approx R_p/(1-e) \approx {3.704} \times 10^{15}\ {\rm cm}$.%, corresponding to an orbital period of $P = 2\pi a^{3/2}/\sqrt{GM_{\rm BH}} \approx \textbf{1423}$ days, \textbf{which is broadly consistent with the inferred recurrence timescale.}

% and an orbital period of
% \begin{equation}
% P \approx 2\pi a^{3/2}/\sqrt{G M_{\rm BH}}
% \approx 1508\ {\rm days}.
% \end{equation}

% The derived timescale is therefore in excellent agreement with the observed recurrence period, providing further support to the partial disruption scenario.

\begin{figure*}[htbp]
    \centering
    
    % Left panel: radio light curves
    \begin{subfigure}{0.48\textwidth}
        \centering
        \includegraphics[width=\linewidth]{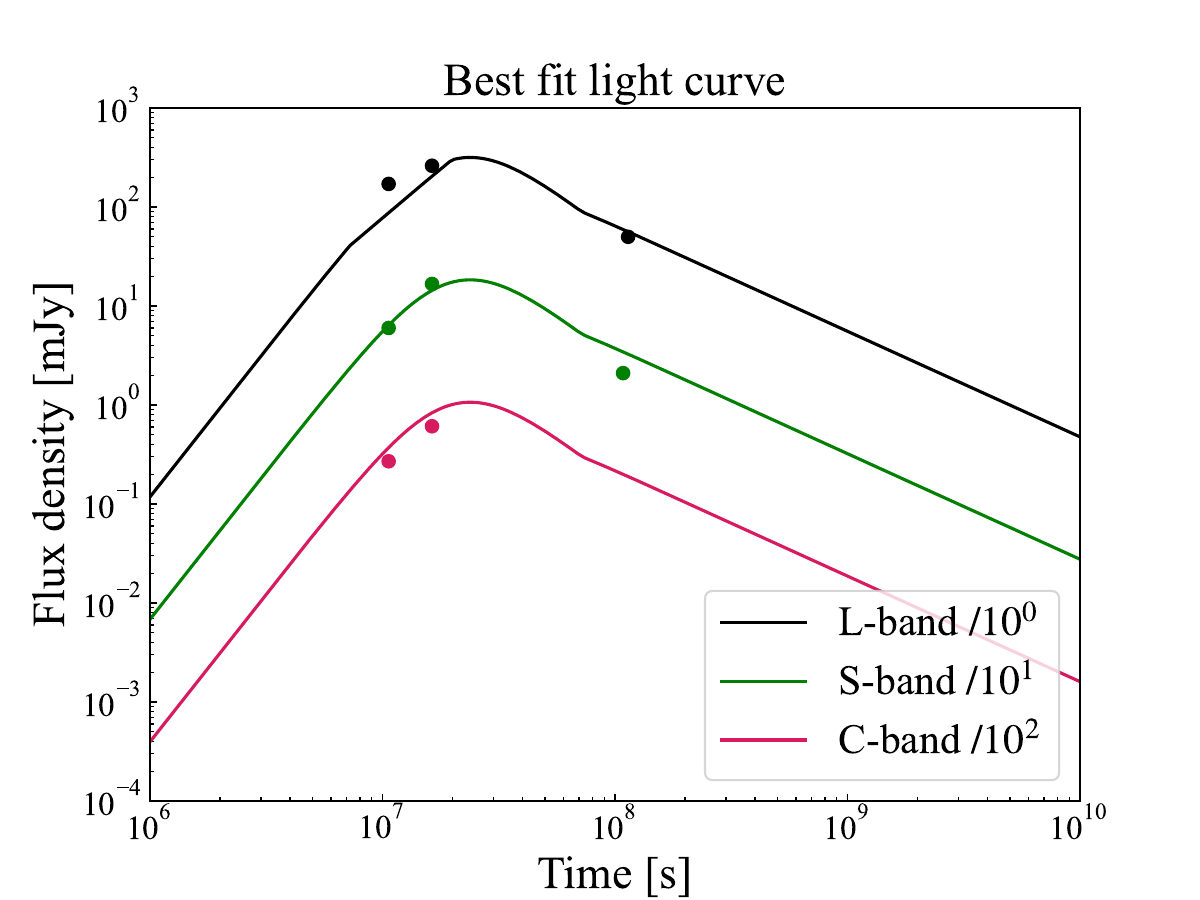}
        \caption{Best fit radio light curves of the late-time emission associated with the second flare.}
        \label{fig:radio-fit}
    \end{subfigure}
    \hfill
    % Right panel: corner plot
    \begin{subfigure}{0.34\textwidth}
        \centering
        \includegraphics[width=\linewidth]{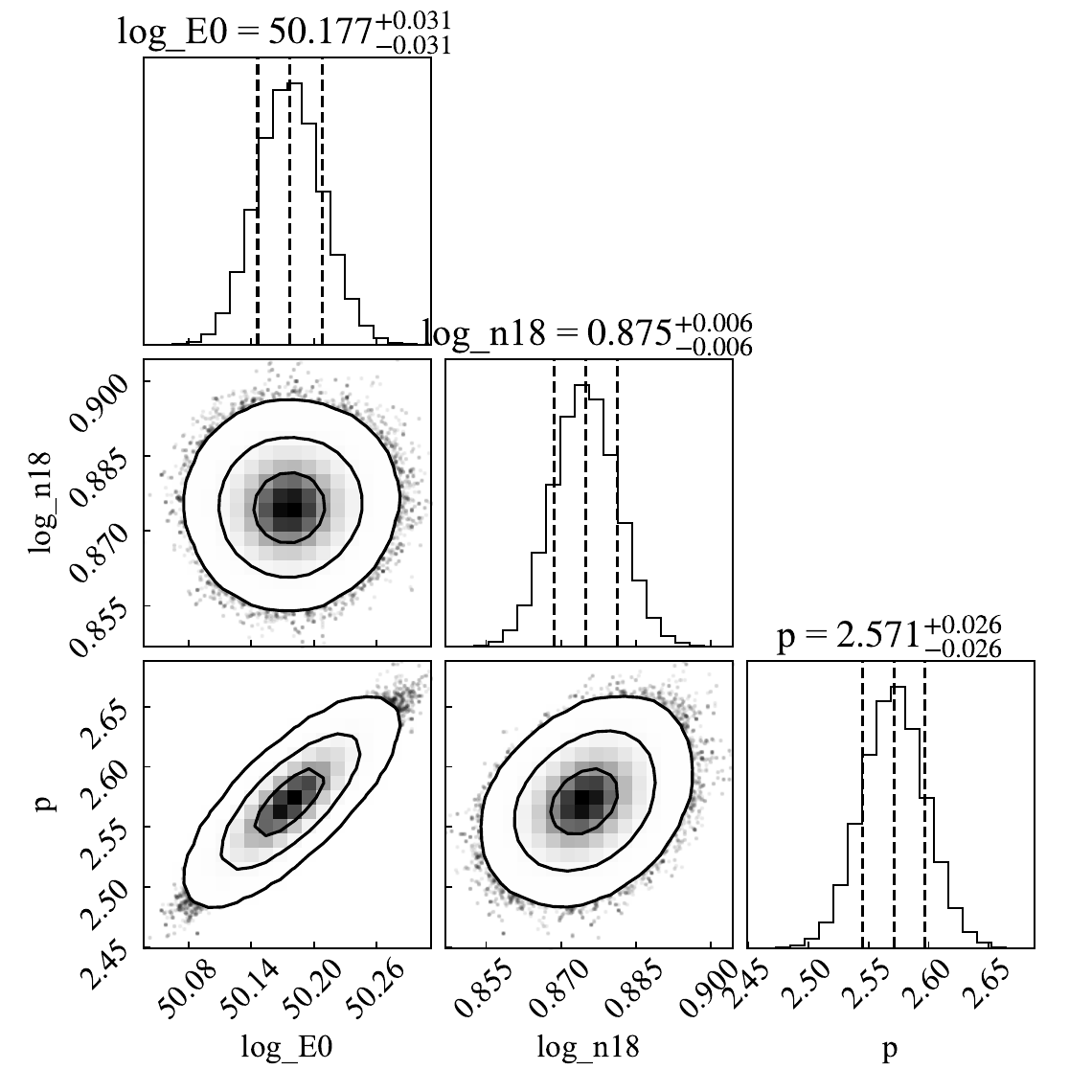}
        \caption{Posterior probability distributions from the MCMC fitting for the second radio flare.}
        \label{fig:corner_plot}
    \end{subfigure}

    \caption{
    Modeling results for the second radio flare of IGR J12580+0134.
    The left panel shows the best-fit radio light curves at 1.5, 3, and 6\,GHz overlaid on the observational data from \citet{Perlman2022}, where the flux densities at 3 and 6\,GHz are scaled down by factors of $10$ and $10^{2}$ for clarity.
    The right panel presents the posterior distributions of the synchrotron afterglow parameters derived from MCMC sampling. %, including the isotropic-equivalent kinetic energy $E_0$, ambient density $n_{18}$, electron spectral index $p$, and microphysical parameters $\epsilon_e$ and $\epsilon_B$.
    }

    \label{fig:radio_modeling}
\end{figure*}

\subsection{Estimating the Disrupted Mass at the Second Encounter}

We adopt the isotropic-equivalent kinetic energy constrained by our
radio afterglow modeling, 
which corresponds to $E_{\rm iso}
\simeq {1.501}\times 10^{50}\ {\rm erg}$.

% Treating the outflow as a two-sided outflow with half-opening angle $\theta_{\rm f}$,
% the beaming correction reads $E_{\rm true} = f_bE_{\rm iso}\simeq 1.50\times 10^{50}\ {\rm erg}$. 
% Here $f_{\rm b} = (1-\cos\theta_{\rm f})$ is the beaming factor, and we adopt $\theta_{\rm f} = 60^\circ$ in our fit.

If we assume a sub-relativistic outflow with a characteristic velocity ${v=0.3c}$ (as fixed in our MCMC analysis), the total kinetic energy of the outflow satisfies $E_{\rm true} \simeq 0.5M_{\rm out}v^2 \sim E_{\rm iso}$,
where $M_{\rm out}$ denotes the total mass participating in the outflow on
both sides. Solving the expression for $M_{\rm out}$ results in ${1.87\times10^{-3}}\,M_\odot$.

This value represents the outflowing mass required to supply the
two-sided kinetic energy inferred from the radio afterglow, under the
assumption that all of this kinetic energy is carried by material moving
at ${v=0.3c}$.

The outflow mass inferred above represents only the material that 
effectively participates in driving the forward shock. 
If the total disrupted mass associated with the second partial stripping
event is denoted by $M_2$, we have
%we introduce two physically motivated fractions:

% \begin{itemize}
% \item $f_M \leq 1$, the mass-loading fraction, defined such that only a fraction $f_M$ of the disrupted mass is launched into the outflow that powers the forward shock;

% \item $\xi \leq 1$, the shock-coupling efficiency, defined as the fraction of the outflow kinetic energy that is converted into the forward-shock energy inferred from radio modeling.

% \end{itemize}
%\left(\frac{f_b}{0.5}\right)
%These definitions imply that
\begin{equation}
\begin{aligned}
M_2  = \frac{2E_{\rm {iso}}}{\xi f_M v^2} \simeq\;& {1.24\times10^{-3}}\,M_\odot
\left(\frac{0.3c}{v}\right)^2
\\[4pt]
&\times
\left(\frac{E_{\rm iso}}{10^{50}\,{\rm erg}}\right)
\left(\frac{1}{f_M}\right)
\left(\frac{1}{\xi}\right) .
\end{aligned}
\end{equation}
% \begin{equation}
% M_2 = \frac{2E_{\rm true}}{\xi f_M v^2},
% \end{equation}
where $f_M$ is the fraction of the disrupted mass that is launched into the outflow, and $\xi$ is the fraction of the outflow kinetic energy that is converted into the forward-shock energy inferred from radio modeling.
% Using the numerical normalization obtained above,we obtain the convenient scaling form

% \begin{equation}
% \begin{aligned}
% M_2 \simeq\;& 0.062\,M_\odot
% \left(\frac{f_b}{0.5}\right)
% \left(\frac{0.03c}{v}\right)^2
% \\[4pt]
% &\times
% \left(\frac{E_{\rm iso}}{10^{50}\,{\rm erg}}\right)
% \left(\frac{1}{f_M}\right)
% \left(\frac{1}{\xi}\right) .
% \end{aligned}
% \end{equation}

We note that the mass scale $M_2$ inferred from the radio afterglow energetics in this section characterizes the amount of material participating in the outflow/fallback during the second passage.

\section{Results and Discussion} \label{sec:highlight}

\subsection{Best-fit Modeling of the second Radio flare}

To investigate the physical origin of the radio emission from IGR J12580+0134, we performed broadband afterglow modeling based on the standard GRB afterglow framework, utilizing the \texttt{PyFRS}\footnote{\url{https://github.com/leiwh/PyFRS}} code \citep{Lei_2016,Zhu2023,Zhou2024,Fu2024}. The modeling was implemented using Markov Chain Monte Carlo (MCMC) techniques to constrain the relevant microphysical and dynamical parameters.

% We adopted MJD 57340 as the zero-point for time calibration, corresponding to the presumed onset of the second radio flare. Due to the limited number of data points during this episode, we fixed several model parameters to reduce degeneracies in the fitting. Specifically, we assumed a uniform interstellar medium (ISM) environment, with the scale index of the ambient density profile fixed at $k = 0$, indicating a constant density environment. The outflow was assumed to be \textbf{sub}-relativistic with a velocity of $\mathbf{v = 0.3c}$, consistent with the expected expansion speed of a mildly energetic TDE ejecta.
We adopt MJD 57340 as the temporal zero-point, corresponding to the presumed onset of the second radio flare. 
%Given the sparse sampling of this episode and the limited spectral information available, not all model parameters can be independently constrained. To reduce parameter degeneracies and avoid over-interpreting poorly constrained microphysical parameters, 
The limited radio data in the second episode are insufficient to constrain all model parameters. We thus fix several quantities in the fitting. We assumed a uniform surrounding CNM. We adopt a sub-relativistic outflow velocity of $v=0.3c$ \citep{piro2025latetimeevolutioninstabilitiestidal}. The microphysical parameters are fixed as $\epsilon_{\rm e}=0.2$ and $\epsilon_{\rm B}=0.1$, for a typical case. Other quantities, i.e., isotropic energy $E_0$, medium density $n_{18}$, and electron spectral index $p$ are treated as free parameters in the fit.
%This allows isotropic energy $E_0$, medium density $n_{18}$, and spectral index of electron $p$ to be treated as free parameters.}  , i.e., the ambient density profile index is fix as $k=0$. The limited radio data are insufficient to constrain the microphysical parameters $\epsilon_{\rm e}$ and $\epsilon_{\rm B}$,

%In addition, because the available radio data provide insufficient leverage to meaningfully constrain

% We also assumed an outflow half-opening angle of $\theta_{\rm f} = 60^\circ$, and an on-axis viewing geometry ($\theta_{\rm obs} = 0^\circ$), such that the line of sight falls within the ejecta cone. These assumptions are justified by the relatively smooth light curves and the lack of evidence for strong beaming effects. 

%, overplotted with the observed data points

Figure~\ref{fig:radio-fit} shows the best-fit light curves at 1.5\,GHz, 3\,GHz, and 6\,GHz. Figure~\ref{fig:corner_plot} is the corresponding corner plot of the posterior distributions. The model successfully reproduces the rise and decay features of the flare across all three bands. 

We note that our modeling of the second radio flare shares important similarities with the approach adopted by \citet{Lei_2016}, while leading to a fundamentally different dynamical interpretation. In both studies, the radio emission is modeled within the standard GRB afterglow framework, where the synchrotron radiation arises from the interaction between an outflow and a CNM. The broadband radio light curves are interpreted using the same underlying synchrotron shock physics.

However, the dynamical properties inferred for the two episodes differ significantly. \citet{Lei_2016} and \citet{yuan2016catching} interpreted the early radio emission following the 2011 TDE as originating from an off-axis relativistic jet interacting with the surrounding medium. In their model, a relativistic outflow with an initial Lorentz factor $\Gamma_0 \gtrsim \mathrm{a~few}$ and a viewing angle $\theta_{\rm obs} \gtrsim 30^\circ$ was required to simultaneously reproduce the radio light curves and satisfy the X-ray constraints.

In contrast, the second radio flare investigated here occurs approximately four years after the first event and is not accompanied by a luminous X-ray counterpart. Our MCMC modeling indicates that the radio emission can be consistently explained by a sub-relativistic outflow with a characteristic velocity $v \approx {0.3c}$ expanding into a uniform ISM environment. No strong beaming correction is required in this case.

This contrast suggests that the two flares may correspond to physically distinct dynamical phases. The first episode likely involved a relativistic jet launched during the initial disruption, whereas the second flare may instead trace a renewed, lower-energy outflow episode associated with a subsequent partial TDE.

\begin{figure}[htbp]
    \centering
    \includegraphics[width=0.48\textwidth]{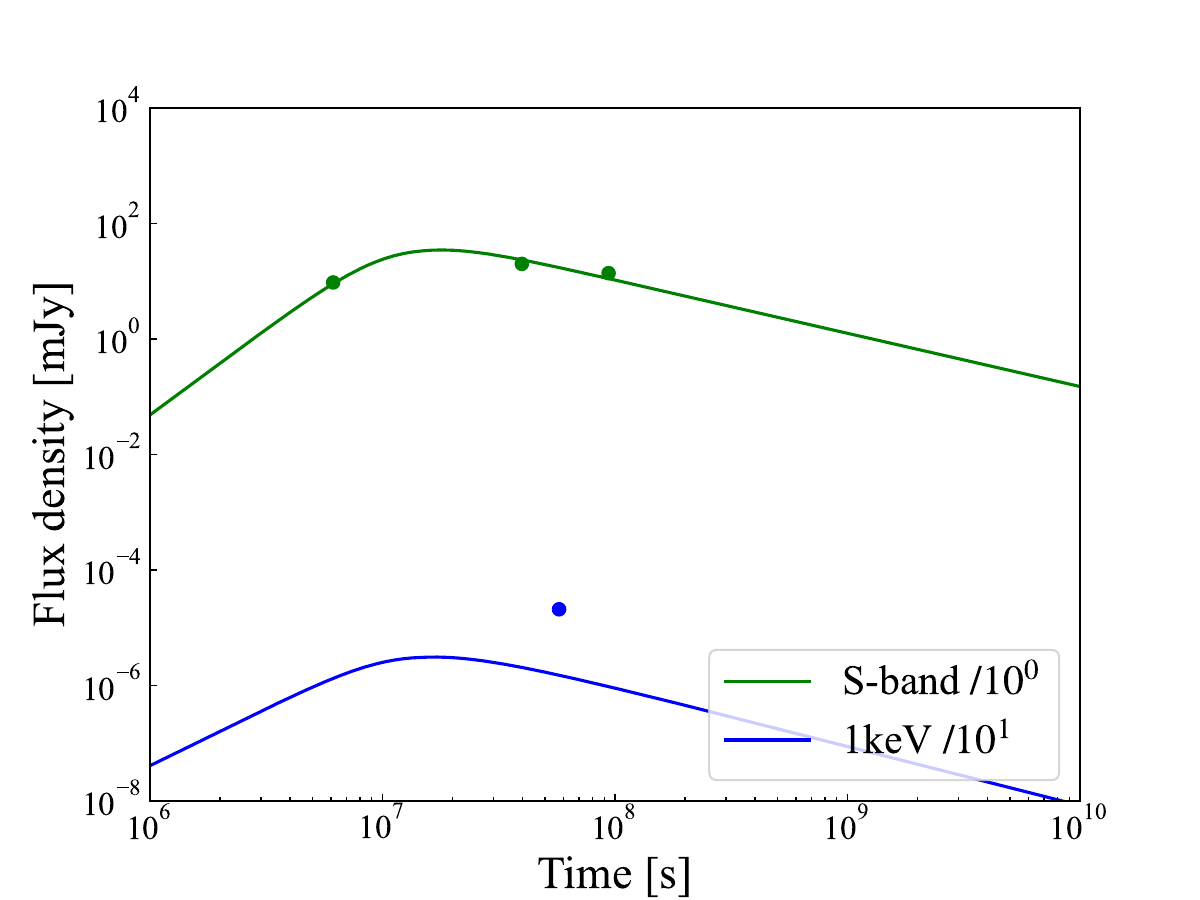}
    \caption{Best-fit light curves for the possible third flare of IGR J12580+0134. 
    The S-band radio data and best-fit model are shown in green, while the 1 keV data point and the corresponding model prediction are shown in blue and scaled by a factor of $10^{-1}$ for visual clarity. The late-time rebrightening can be interpreted as synchrotron emission from an outflow interacting with the CNM.
}
\label{fig:3_flare}
\end{figure}

We also performed a  fit to the possible third radio flare, as shown in Figure \ref{fig:3_flare}.
Due to poor data, the onset time of this episode cannot be robustly constrained.
Therefore, we adopt MJD~59400 as the starting time for such newly launched wind, in the fitting procedure.
The best-fit parameters are 
$E_0 = 2.62 \times 10^{49}\ {\rm erg}$, 
$n_{18} = 7.24\ {\rm cm^{-3}}$, 
$\epsilon_e = 0.25$, 
$\epsilon_B = 0.068$, 
and $p = 2.09$.
The remaining dynamical parameters are fixed to the same values as those adopted for the second radio flare (i.e., $v=0.3c$).

\subsection{Is dust extinction responsible for the optical faintness?}

\begin{figure}[htbp]
    \centering
    \includegraphics[width=0.48\textwidth]{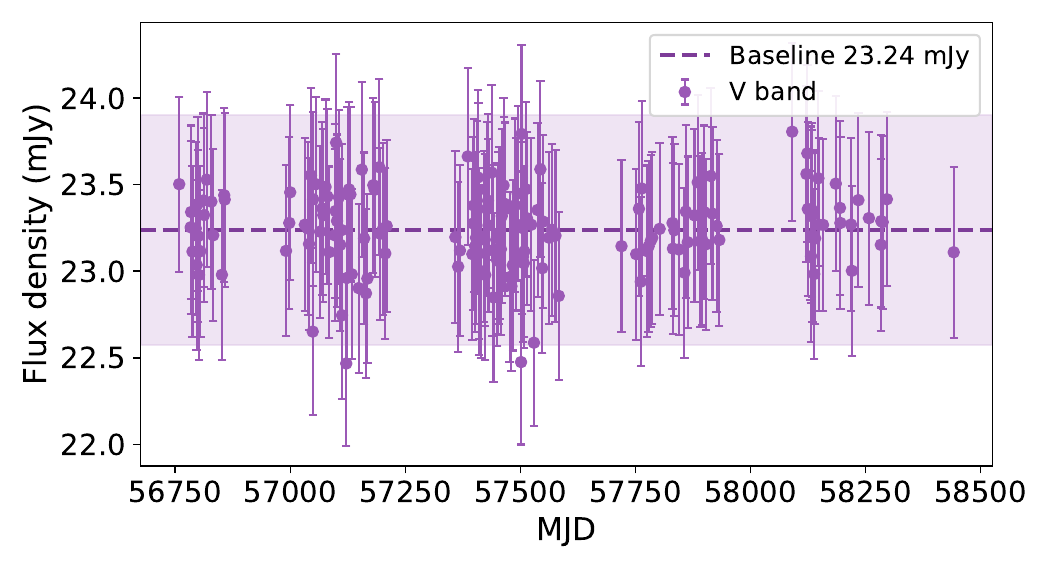}
    \caption{Optical light curve of IGR~J12580+0134 in the ASAS-SN $V$ band.
    The purple points show the ASAS-SN $V$-band flux density as a function of MJD.
    The dashed horizontal line indicates the weighted mean baseline flux 
    ($F_0 = 23.24\,\mathrm{mJy}$), and the shaded region represents the $\pm3\sigma$ 
    dispersion around this baseline.
    No significant optical variability beyond the $3\sigma$ level is detected 
    during the second radio flare.
}
\label{fig:optical}
\end{figure}

As shown in Figure \ref{fig:optical}, no significant optical variability is detected during the second radio flare, with all measurements remaining within the 3$\sigma$ level. These data were obtained from the ASAS-SN survey, specifically in the V-band, which provides optical monitoring of transient events.
From the observations we can compute a upper limit 
on the optical variability.
The derived observed upper limit of $\nu L_\nu (V) \lesssim 1.4 \times 10^{41}\ \mathrm{erg\,s^{-1}}$. We estimate the dust extinction implied by the intrinsic
hydrogen column density. Since no constraint on $N_{\rm H}$ is available for the second radio flare,
we adopt the value derived from the 2023 X-ray flare for our extinction estimate. 
We estimate the optical extinction using the empirical relation
$N_{\rm H}/A_V \simeq 1.79\times10^{21}\ \mathrm{cm^{-2}\,mag^{-1}}$ \citep{Predehl1995A&A...293..889P}. For $N_{\rm H}=8\times10^{20}\ \mathrm{cm^{-2}}$ \citep{Danehkar_2025},
this gives $A_V \approx 0.45$ mag. Assuming a standard Galactic extinction law 
\citep{Cardelli1989ApJ...345..245C} with $R_V=3.1$, the inferred $A_V \approx 0.45$ mag implies that the 
extinction-corrected emission (or variability upper limit) in the ASAS-SN $V$ band can be 
at most a factor of $10^{0.4A_V} \approx 1.51$ higher.
Applying this correction increases from $\nu L_\nu(V)\lesssim 1.4\times10^{41}$ to
$\nu L_\nu(V)\lesssim 2.1\times10^{41}\ \mathrm{erg\,s^{-1}}$. This modest correction indicates that dust extinction cannot account for the optical faintness, 
suggesting that the weak optical emission during the second flare of IGR J12580+0134 is an intrinsic characteristic of the event.

We note that the extinction estimate above is based on the column density measured during
the 2023 X-ray flare ($N_{\rm H}\simeq 8\times10^{20}\ \mathrm{cm^{-2}}$), which is nearly two
orders of magnitude lower than the value reported for the first X-ray flare,
$N_{\rm H}\sim 7\times10^{22}\ \mathrm{cm^{-2}}$ \citep{nikolajuk2013tidal}. As an illustrative scenario, if the line-of-sight column during the second radio flare were
substantially higher than our fiducial value---e.g., $N_{\rm H}\approx 7\times10^{21}\ \mathrm{cm^{-2}}$---and if the gas-to-dust ratio were close to the Galactic value, the implied extinction would be
$A_V \approx 3.9$ mag, corresponding to a flux correction factor of $\sim 36$.
In this case, the extinction-corrected optical luminosity would increase to
$\sim 5\times10^{42}\ \mathrm{erg\ s^{-1}}$, consistent with the faint end of optical TDEs.
Such a low optical luminosity would be naturally explained if the disrupted object were less massive
than in typical optically bright TDEs \citep{Velzen2020SSRv..216..124V}, although this inference remains model dependent.

\subsection{Possible Explanations for the Lack of an X-ray Counterpart to the Second Flare}

The lack of a statistically significant X-ray brightening during the second radio flare does not necessarily imply the absence of accretion activity. Within a repeated pTDE scenario, the 2016 event need not produce an X-ray flare comparable to the 2011 event. The amount of stellar material stripped during each encounter may differ substantially, leading to a different fallback rate and hence a different accretion-powered X-ray luminosity. In this picture, the second partial disruption could have been intrinsically weaker, shorter-lived, and/or softer in X-rays, while still generating an outflow whose interaction with the CNM produced the observed radio rebrightening. Swift/XRT observations obtained around the epoch of the radio rebrightening had limited photon counts, requiring the co-addition of multiple datasets to achieve an adequate signal-to-noise ratio. Sparse temporal coverage would therefore make such an intrinsically faint or short-lived X-ray episode even more difficult to detect.

One natural possibility is that the accretion-powered X-ray peak occurred before the available X-ray observations. In TDEs, X-ray radiation is generally expected to trace the fallback and early accretion phases, whereas radio emission typically arises from the delayed interaction between the outflow and the surrounding medium \citep{Rees1988}. If the second partial disruption followed a similar evolutionary sequence, the onset of renewed accretion could have preceded the observed rise of the radio flare, such that the X-ray peak was missed by the sparse temporal coverage. This timing effect would be especially important if the second X-ray episode was already weaker or shorter-lived than the 2011 outburst.

Obscuration may also play a secondary role. The first flare exhibited a heavily absorbed spectrum with a hydrogen column density of $N_{\rm H} \sim 7\times10^{22} \mathrm{cm^{-2}}$ \citep{nikolajuk2013tidal}, demonstrating that substantial line-of-sight material can exist in this system while still allowing detectable hard X-ray emission. However, obscuration alone is unlikely to fully account for the non-detection unless the intrinsic luminosity of the second event was lower or its spectrum significantly softer than that of the 2011 outburst. Obscuration is therefore interpreted as a secondary effect rather than the primary mechanism responsible for the contrasting X-ray behaviours exhibited by the two episodes. A reduced fallback mass during the second encounter would naturally lead to a weaker accretion flow, consistent with the expectations for repeated partial tidal stripping.

An alternative possibility is that the second episode was dominated by kinetic energy carried by the outflow rather than by radiatively efficient accretion. In this scenario, the radio emission would primarily trace the shock interaction with a dense circumnuclear environment, while the high-energy signature could remain intrinsically faint \citep{alexander2020radio}. Such behavior has been suggested in several low-luminosity or inefficiently accreting TDEs. While the current data do not uniquely require this interpretation, it remains a plausible contributor to the weak high-energy signature during the second episode.

Overall, the contrasting X-ray behaviors of the two episodes do not rule out renewed accretion during the second event. Rather, they can be naturally understood if the second passage stripped less material and therefore produced a weaker, shorter-lived, and/or softer accretion-powered X-ray flare, with sparse temporal coverage and possible moderate obscuration further reducing its detectability. Deeper and higher-cadence X-ray monitoring will be essential for determining whether future passages produce detectable high-energy counterparts, thereby clarifying the accretion properties of this candidate repeating partial tidal disruption system.

\subsection{Alternative Interpretations of the Late-time Radio Rebrightening}
\label{sec:alt_scenarios}

% \textbf{Before interpreting the 2016 radio flare in IGR~J12580+0134 as evidence for a repeated pTDE, it is important to distinguish between two separate questions: (1) what physical process directly powers the observed late-time radio emission, and (2) what triggers that process. This distinction is especially important because the observational situation in IGR J12580+0134 differs from that in the well-studied repeated pTDE systems (e.g. AT2022dbl, AT2020vdq). By contrast, in IGR J12580+0134 the repeated activity is presently seen only in the radio band, while no comparably luminous second X-ray flare has been detected. We therefore do not regard the current multi-wavelength data as directly demonstrating a repeated pTDE in the same sense as AT2022dbl or AT2020vdq. Instead, the question is whether the late-time radio rebrightening is better understood as (i) interaction of the original outflow with dense clouds, (ii)  multiple outflows associated with a change in the accretion state, or (iii) emission from a relativistic jet viewed off-axis.}

Unlike AT2022dbl and AT2020vdq, the repeated activity of IGR J12580+0134 is currently observed only in the radio band, while no comparably luminous second X-ray flare has been detected. We therefore do not regard the current multi-wavelength data as conclusive evidence of a repeated pTDE analogous to AT2022dbl or AT2020vdq. Instead, three alternative mechanisms could explain the multi-wavelength observations: (i) interaction between the original outflow with dense clouds, (ii)  multiple outflows associated with a change in the accretion state, or (iii) emission from a relativistic jet viewed off-axis.

\begin{figure}[htbp]
    \centering
    \includegraphics[width=0.48\textwidth]{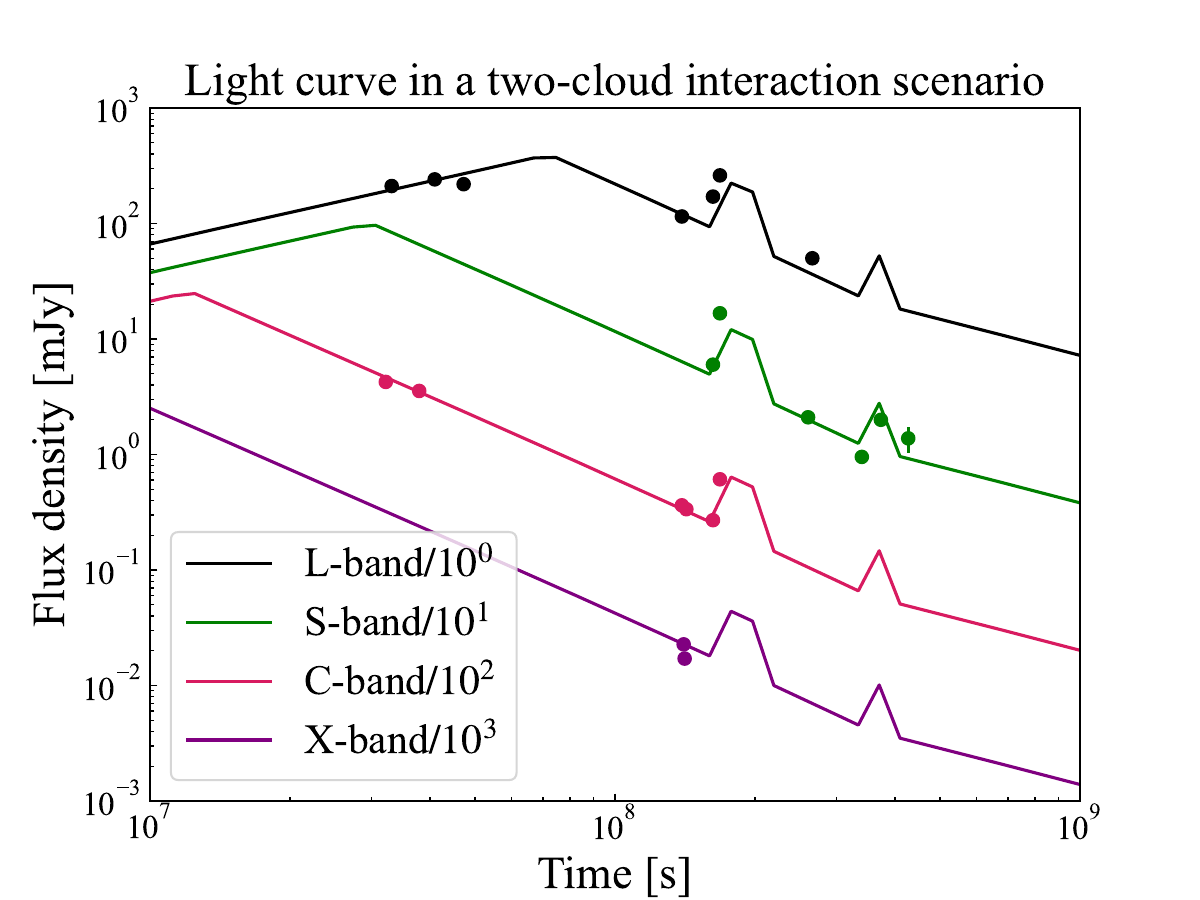}
    \caption{Radio light curves in a two-cloud interaction scenario. Circles denote the observed flux densities in the L-, S-, C-, and X- bands, while the solid lines show the corresponding model light curves. The flux densities in the S-, C-, and X-bands have been scaled by factors of $10$, $10^2$, and $10^3$, respectively, for clarity.
}
\label{fig:density}
\end{figure}

The first possibility is that the radio rebrightening was caused by interaction between the original outflow with dense clouds\citep{Perlman2022,yang2025outflowcloudinteractionpossibleorigin,lei2024debrisstreamtorus,zhuang2025cloudradioflare}. This is closely related to the interpretation previously proposed for IGR J12580+0134, in which the original jet eventually encountered an overdense cloud and produced a second radio flare \citep{Perlman2022}. 
We also explored a two-cloud version of this picture, illustrated in Figure~\ref{fig:density}, to assess whether a second cloud encounter could explain the putative third radio flare. Our results suggest that this model can account for the general characteristics of the radio light curve. Nevertheless, if the putative third radio flare is physically associated with the weak X-ray activity seen in 2023, then such a radio/X-ray association would be less naturally explained within the simple cloud-interaction picture of \citet{Perlman2022}, in which no prominent X-ray flare was expected or observed during the 2016 radio rebrightening. We therefore regard dense-cloud model as a plausible contributor to the radio light curve's profile, but not as the most compelling explanation for the long-term evolution of the source.

% \textbf{A further caveat is that the source may also show a later episode of weak reactivation. In particular, the S-band light curve may hint at an additional late-time excess, and NICER monitoring revealed faint X-ray flares in 2023. The timescale of this putative later activity is also broadly comparable to the separation between the first and second radio flares. If this feature is real and physically associated with a new radio-brightening episode, then such a roughly repeated temporal spacing would make a simple one-time interaction between the original outflow and a single overdense cloud less natural, unless the CNM contains multiple overdense structures or strong radial inhomogeneity. At present, however, the existence and interpretation of such a third radio episode remain uncertain, so we do not treat this as a decisive argument against the density-enhancement scenario.}

The second possibility to account for late-time radio flares is the presence of multiple outflows. A representative example is ASASSN-15oi, for which \citet{Horesh_2021} reported a radio flare detected half a year after discovery and a second, brighter flare years later. They argued that a standard picture in which the outflow is launched promptly at disruption cannot easily account for the combined temporal and spectral behavior, and instead suggested delayed ejection, possibly associated with an accretion-state transition \citep{Horesh_2021}. In AT2019azh, \citet{Sfaradi_2022} found that the late-time radio flare began roughly contemporaneously with a dramatic X-ray brightening and softening, and proposed that the radio/X-ray evolution may reflect a change in accretion state, paralleling the state transitions in X-ray binaries. 

% In AT2020vwl, \citet{Goodwin2024} identified a second radio flare and argued for two distinct ejection episodes, or renewed energy injection into a pre-existing outflow; importantly, they showed that the rapid decay of the second flare is not easily explained by environmental changes alone, while the inferred outflow speed is too low for a simple off-axis relativistic-jet interpretation. More recently, \citet{Sfaradi2025} considered a broader set of possibilities---including prompt outflows in a structured medium, delayed sub-relativistic outflows, prompt off-axis jets, and late-time jets---to interpret the rapidly double-peaked radio emission of AT2024tvd.

The theoretical motivation for late-time outflow launching from long-lived TDE accretion disks has also strengthened in recent years. \citet{piro2025latetimeevolutioninstabilitiestidal} showed that thermal instabilities can drive repeated transitions between high and low accretion states for up to years, with the super-Eddington high states potentially associated with mass ejections of ${\sim10^{-3}-10^{-1}} M_\odot$ and velocities of ${\sim0.03-0.3c}$, which may contribute to late-time radio flares. Building upon this framework, \citet{Wu_2026} modeled collisions involving these delayed disk-driven outflows—both among themselves and with the CNM, showing that the resulting shocks can reproduce radio luminosities and characteristic velocities comparable to those inferred in delayed radio TDEs.

% \textbf{In this context, the repeated pTDE should be viewed not as a separate radio emission mechanism, but rather as one possible trigger for such renewed activity. However, the observational distinction between renewed outflows from a repeated pTDE and generic multiple outflows due to changes in accretion state cannot rest on recurrence alone \citep{piro2025latetimeevolutioninstabilitiestidal,Wu_2026}.  
% A stronger case for a repeated pTDE instead comes when successive episodes resemble new TDE-like thermal flares with closely similar near-peak photometric and spectroscopic evolution, such as AT 2022dbl and AT 2020vdq \citep{Lin_2024,Somalwar2025-2020vdq}. By contrast, current multiple outflows models mainly address the late-time radio phenomenology, and it remains unclear whether they would naturally reproduce nearly identical multi-epoch line evolution across separate episodes. }

For IGR J12580+0134, both a repeated pTDE and accretion state transitions can generate relativistic outflows. We emphasize, however, that the distinction between renewed outflows driven by a repeated pTDE and multiple outflows associated with accretion-state changes cannot be established from recurrence alone. \citet{piro2025latetimeevolutioninstabilitiestidal} and \citet{Wu_2026} show that repeated late-time activity can arise in long-lived TDE accretion systems without invoking a repeated pTDE. In the present source, however, the first two major activity episodes, together with a possible third radio rebrightening, may hint at a characteristic recurrence timescale, which is not generically implied by multiple-outflow scenarios alone.

The third possibility is that the second radio flare originates from an off-axis relativistic jet, a scenario that has gained significant prominence in recent literature on delayed-radio TDEs. \citet{Beniamini_2023} argued, in the context of Swift J1644+57, that slightly off-axis viewing can produce an apparent increase in the inferred jet energy, as the jet decelerates and progressively larger portions of it come into view. Based on closure relation analysis of spectral and temporal flux evolution, they showed that the source aligns with an off-axis jet model. \citet{Matsumoto_2023} generalized equipartition analysis to relativistic off-axis ``top-hat" jets and showed that, as such a jet decelerates, an off-axis observer can see a progressively larger fraction of the emitting region, leading to an apparent increase in the inferred energy and a rapidly rising radio light curve. In the context of TDEs, they suggested that the delayed radio flare of AT2018hyz could be consistent with a relativistic off-axis jet. Building on this framework, \citet{Sfaradi2024MNRAS.527.7672S} modeled AT2018hyz with a narrow, powerful jet viewed at a relatively large angle from the jet axis. Most recently, continued monitoring of AT2018hyz by \citet{cendes2025continuedrapidradiobrightening} showed that the late-time radio evolution remains compatible with either a delayed spherical outflow or a highly off-axis relativistic jet, highlighting that these two interpretations remain difficult to distinguish without further constraints from continued light-curve evolution or VLBI observations. A different but related interpretation was proposed by \citet{teboul2023unifiedtheoryjettedtidal}, who argued that delayed radio flares may arise in initially misaligned TDE disk--jet systems, where the jet launch only after the system has aligned more closely with the black hole spin. Within this unified framework, promptly launching relativistic jets and delayed radio flares are treated as different manifestations of the same underlying TDE-jet phenomenology. 
\begin{figure}[htbp]
    \centering
    \includegraphics[width=0.48\textwidth]{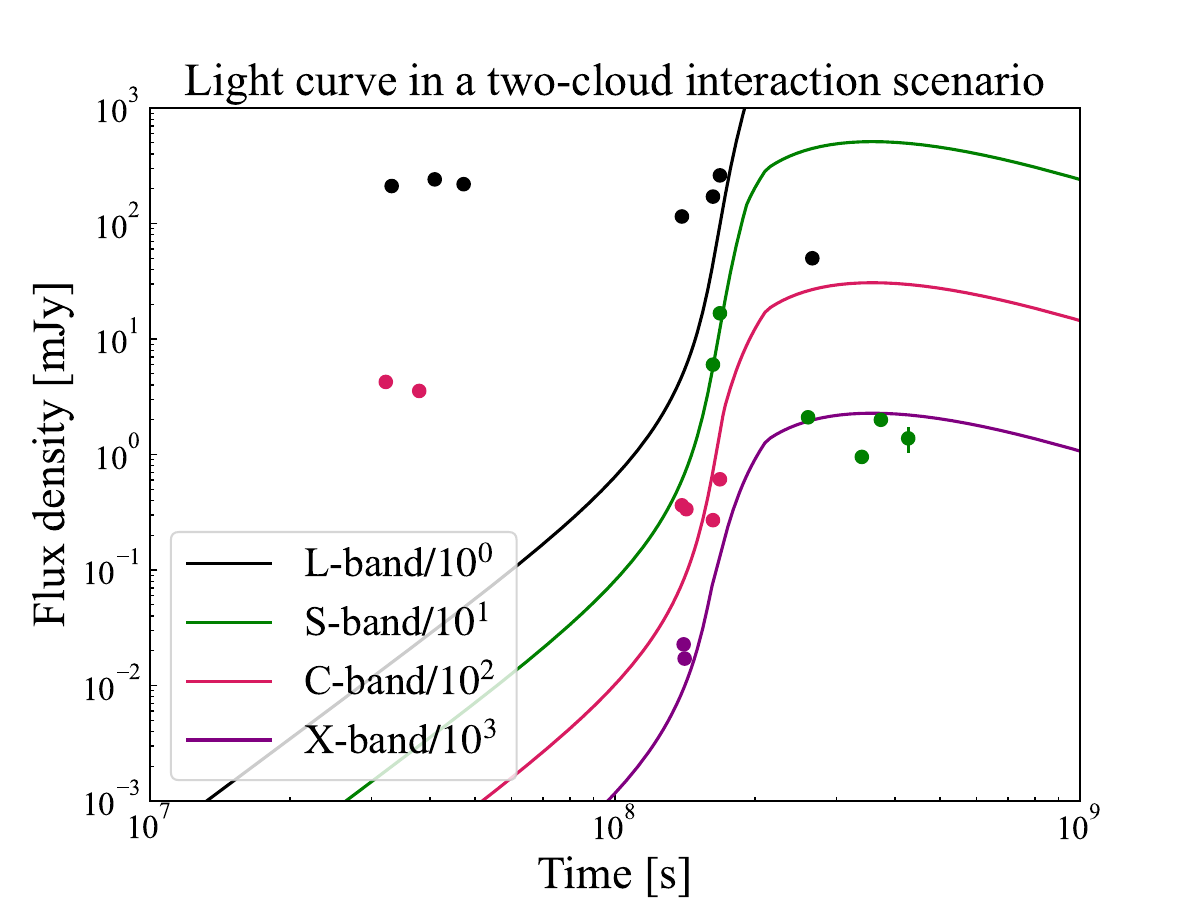}
    \caption{Radio light curves of the second radio flare in IGR J12580+0134, shown together with a representative fit in an off-axis scenario. Dots with error bar denote the observed flux densities in the L-, S-, C-, and X- bands, while the solid lines show the corresponding model light curves. The flux densities in the S-, C-, and X-bands have been scaled by factors of $10$, $10^2$, and $10^3$, respectively, for clarity. The dashed line marks the power-law fit to the late-time decay of the S-band data.}

\label{fig:off-axis}
\end{figure}

We also explored whether the second radio flare could be reproduced in an off-axis jet scenario. A representative fit is shown in Figure~\ref{fig:off-axis}, obtained with $E_0=6.27\times10^{54}\,\mathrm{erg}$, $\Gamma_0=13.25$, $\theta_{\rm j}=11.45^\circ$, $\theta_{\rm obs}=41.79^\circ$, $\epsilon_{\rm e}=0.27$, $\epsilon_{\rm B}=0.09$, $p=2.47$, and $n_{18}=2.45\times10^{-5}\,\mathrm{cm^{-3}}$. As can be seen, the off-axis model can reproduce the rise of the second radio flare reasonably well, but it fails to provide a satisfactory description of the full radio evolution. In particular, the off-axis model does not provide a satisfactory description of the radio data associated with the first flare, because it predicts a low early-time flux, in contrast to the relatively bright radio emission observed at early times. Moreover, the late-time decline of the S-band data is not well captured by the model. A power-law fit to the S-band decay gives $F_\nu \propto t^{-4.29}$, implying a rather steep fading rate that is difficult to reconcile with a simple off-axis interpretation. We therefore conclude that, although an off-axis configuration can qualitatively account for part of the rebrightening behavior, it does not provide a satisfactory overall explanation for the second radio flare observed in IGR J12580+0134.

Overall, while the alternative scenarios may account for parts of the observed behavior, the repeated pTDE scenario provides the most self-consistent interpretation of the second radio flare, as well as the possible third radio rebrightening and its weak X-ray activity in 2023, in IGR J12580+0134.

\section{Summary} \label{sec:sum}
We present a multi-wavelength study of IGR J12580+0134 in NGC 4845, motivated by the late-time radio rebrightening reported several years after the 2011 X-ray tidal-disruption flare \citep{nikolajuk2013tidal,Irwin2015,Perlman2022}. We compile the available radio and X-ray measurements and model the second radio flare within a standard synchrotron afterglow framework.

Our radio modeling constrains the outflow energetics and ambient density associated with the 2016 episode, indicating a mildly energetic, sub-relativistic ejecta interacting with a dense circumnuclear environment. The inferred energetics imply that only a modest amount of mass is required to power the radio emission, consistent with a partial mass-loss episode rather than a single complete disruption. In this context, the comparatively low disrupted mass inferred for the 2011 event can be interpreted as the stripped mass that entered fallback and accretion, rather than the total mass of the disrupted object \citep{nikolajuk2013tidal,Lei_2016}.

On the high-energy side, Swift/XRT observations obtained during the 2016 radio flare did not show a statistically significant X-ray brightening \citep{Perlman2022}. More recently, NICER monitoring revealed faint X-ray flares during March--June 2023 that are substantially weaker than the 2011 outburst, suggesting intermittent low-level accretion activity in the nucleus \citep{Danehkar_2025}.

We also examined several possible interpretations for the late-time radio rebrightening. Although the current data do not allow a unique identification of the underlying mechanism, the repeated pTDE scenario provides a relatively natural overall explanation for the observed multi-epoch radio/X-ray behavior.

Taken together, our results support a scenario in which IGR J12580+0134 is a promising candidate for a repeating pTDE. Future high-cadence, sensitive X-ray monitoring and continued radio follow-up will be essential for testing this interpretation and for searching for additional episodes on multi-year timescales.

\begin{acknowledgments}
We thank the participants of the TDE FORUM (Full-process Orbital to Radiative Unified Modeling) online seminar series for their inspiring discussions. We are very grateful to Rongfeng Shen, Erlin Qiao, and Chichuan Jin for their helpful discussions. This work is supported by the National Key R\&D Program of China (No. 2023YFC2205901) and the National Natural Science Foundation of China under grants 12473012, 12533005, and 12321003, the Fundamental Research 
Funds for the Central Universities, HUST (No. YCJJ20252115). This work is supported by Project for Young Scientists in Basic Research of Chinese Academy of Sciences (No. YSBR-061). 

%W.H.Lei. acknowledges support by the science research grants from the China Manned Space Project with NO.CMS-CSST-2021-B11.

\end{acknowledgments}

\begin{deluxetable*}{lccccc}[htbp]
\caption{List of published repeated pTDE candidates\label{tab:rptde}}
\tablehead{
\colhead{Name} & \colhead{Host Type} & \colhead{Band} & \colhead{Period/Interval (Days)} & \colhead{Flares} & \colhead{Peak Evolution}
}
\startdata
ASASSN-14ko$^{1,2,3,4,5}$ & Seyfert 2 & Opt./UV/X-ray$^{\dagger}$ & 115.2 & $\sim$30 & Similar \\
Swift J023017.0+283603$^{6,7}$ & Weak AGN & X-ray & $\sim$22 & $\sim$11 & Variable\\
eRASSt J045650.3--203750$^{8}$ & Quiescent & X-ray/UV$^{\dagger}$ & 299$\rightarrow$193 & 5 & Lower\\
AT\,2023uqm$^{9}$ & Quiescent & Opt./UV/X-ray$^{\ddagger}$ & $\sim$527 & $\geq$5 & Increasing\\
\tableline
IC 3599$^{10,11,12,13,14}$ & Seyfert 1.9 & X-ray/Opt.$^{*}$ & $\sim$3470$^{\star}$ & 2/3 & Similar \\
RX J133157.6--324319.7$^{15,16}$ & Quiescent & X-ray & $\sim$10\,000 & 2 & Similar\\
AT\,2018fyk$^{17,18,19}$ & Quiescent & UV/X-ray & $\sim$1200 & 2 & Lower \\
AT\,2019aalc$^{20}$ & Seyfert 1 & Opt./UV/X-ray$^{*}$ & $\sim$1460 & 2 & similar \\
AT\,2020vdq$^{21,22}$ & E+A & Opt./UV$^{*}$/X-ray$^{*}$ & $\sim$870 & 2 & Higher\\
AT\,2021aeuk$^{23}$ & Seyfert 1 & Opt./UV/X-ray$^*$ & $\sim$1100 & 2 & Variable\\
AT\,2022dbl$^{24,25,26,27}$ & QBS & Opt./UV & $\sim$710 & 2 & Lower\\
F01004-2237$^{28}$ & Dusty starburst & Opt./UV/X-ray$^*$ & $\sim$3760 & 2 & Lower\\
AT\,2023adr$^{29}$ & Quiescent & Opt./UV & $\sim420$ & 2 & Similar \\
AT\,2022sxl$^{30}$ & Composite & Opt./MIR & $\sim2600$ & 2 & Similar \\
AT~2019azh$^{31}$ & E+A & Opt. & $\sim4800$ & 2 & Similar \\
AT~2024pvu$^{31}$ & Green valley & Opt./UV & $\sim6200$ & 2 & Similar \\
\enddata

\tablecomments{\\
$^{\dagger}$ Not periodic. \\$^{*}$ Not observed during the first flare. \\$^{\ddagger}$ X-ray observations were obtained only during the fifth flare.\\
% Period/Interval: ASASSN-14ko shows a nearly constant period of 115.2 days. Swift J023017.0+283603 shows a period of $\sim$22 days. eRASSt J045650.3--203750 has shown 5 flares with the interval declining from 299 days to $\sim$193 days. Other sources show only two flares. 
$^{\star}$ IC~3599 exhibited two prominent X-ray flares in 1990 and 2010. 
Based on a repeated pTDE interpretation, \citet{Campana2015} proposed a recurrence period of $\sim$9.5~yr, implying a missed flare between the two events. 
However, subsequent X-ray monitoring did not detect an additional flare within the predicted time window \citep{Grupe2024ApJ...969...98G}.
 \\
% Peak Evolution: Peak luminosity of the later flare relative to the earlier flare. \\
References: 1. \citet{payne2021asassn}, 
2. \citet{Payne2022ApJ...926..142P}, 
3. \citet{Payne2023ApJ...951..134P}, 
4. \citet{Huang2023ApJ...956L..46H}, 
5. \citet{bandopadhyay2024repeating}, 
6. \citet{Evans2023NatAs...7.1368E}, 
7. \citet{Guolo2024NatAs...8..347G}, 
8. \citet{Liu2024}, 
9. \citet{wang2025starsdeaththousandcuts}, 
10. \citet{Grupe1995}, 
11. \citet{Komossa1999}, 
12. \citet{Grupe2015ApJ...803L..28G}, 
13. \citet{Campana2015}, 
14. \citet{Grupe2024ApJ...969...98G}, 
15. \citet{Hampel2022RAA....22e5004H}, 
16. \citet{Malyali2023MNRAS.520.3549M}, 
17. \citet{Wevers2019MNRAS.488.4816W}, 
18. \citet{Wevers2023ApJ...942L..33W}, 
19. \citet{pasham2024potentialsecondshutoffat2018fyk}, 
20. \citet{veres2025deadat2019aalccandidaterepeating}, 
21. \citet{Yao2023}, 
22. \citet{Somalwar2025-2020vdq},
23. \citet{sun2025at2021aeuk},
24. \citet{2024TNSAN..43....1Y},
25. \citet{Lin_2024}
26. \citet{hinkle2025doubleluminousflaresnearby},
27. \citet{makrygianni2025doubletidaldisruptionevent}
28. \citet{Sun_2024}
29. \citet{angus2026tidaldisruptioneventmodels}
30. \citet{ji2025at2022sxlcandidaterepeatingtidal}
31.\citet{yao2026raterepeatingtidaldisruption}
}
\end{deluxetable*}

\begin{table*}[ht]
\centering
\caption{VLA Observations}
{\footnotesize
\begin{tabular}{llrcccccccc}
\hline
\hline
Date$^a$& Project$^b$ & $\Delta$ T$^c$ &Config$^d$ &Freq$^e$ (Band)&$\Delta$ B$^f$  & Beam$^g$ & RMS$^h$  & Nuclear Flux$^i$  \\
&&(days)&&(GHz)&(MHz)&($\prime\prime$, $\prime\prime$, $\circ$) & (mJy/beam) & Density (mJy) \\
\hline
1995-Feb-27 & NVSS$^j$ &  -6150 &D &1.4 (L)& 50&  45, 45, 0&0.45&26$\,\pm\,4$\\
1998-Jul-26 & AB0879$^k$ & -4905 & B & 1.4 (L) & 100   &  5.1, 4.2, 24.0& 1.0   & 34$\pm\,2$\\
2011-Dec-19 & 10C-119$^l$ & $-$11& D & 6.0 (C) & 2048  &  11.0, 9.1, -1.4&  0.015  & 425$\,\pm\,3$\\
2011-Dec-30 (T1)& 10C-119$^l$ & 0    & D & 1.6 (L) & 500  &  38.6, 34.3, -5.22&  0.040 & 211$\,\pm\,3$\\
2012-Feb-24 & 10C-119$^l$ &56& C & 6.0 (C) & 2048  &  3.1, 2.8, -11.7&  0.0039 & 355$\,\pm\,2$\\
2012-Mar-30 & 10C-119$^l$ & 91& C & 1.6 (L) & 500  &  12.2, 11.1, -41.8&  0.045 &  241$\,\pm\,3$\\
2012-Jun-11 & 10C-119$^l$ & 164& B & 1.6 (L) & 500  &  3.5, 3.3, 22.7&  0.018 &  219$\,\pm\,4$\\
2015-Jun-22 & 15A-357$^m$ & 1270 & A & 1.5 (L) & 1000 & 1.73, 1.06, 43.2& 0.12& 115$\,\pm$2 &\\
2015-Jun-22 & 15A-357$^m$ &1270&A &5.5 (C)&1000 & 0.47, 0.33, 46.2&  0.013 &36.3$\,\pm\,0.3$\\
2015-Jul-06 & 15A-400$^k$& 1284&A&9 (X)&2048&0.21, 0.2, 20 & 0.010  & $22.7\pm0.3$  \\
2015-Jul-14 & 15A-400$^k$& 1292&A&9 (X)&2048 & 0.27, 0.16, 43 & 0.020  & $17.1\pm0.2$ \\
2015-Jul-28 & 15A-400$^k$ & 1306 &A&6 (C)&2048& 0.37, 0.28, 48.5 & 0.015&  $33.6\pm0.2$ \\
 2016-Mar-16 & 16A-420$^k$ & 1538 &  C & 6.0 (C)& 2048  &3.5, 3.0, -52.5&  0.012  &   27$\,\pm\,2$&\\
 2016-Mar-16 & 16A-420$^k$ & 1538 &  C & 3.0 (S)& 2048  &6.4, 5.2, -18.4&   0.032 &  $60\,\pm\,3$&\\
 2016-Mar-16 & 16A-420$^k$ & 1538 &  C & 1.5 (L)& 1024 &12.9, 9.9, -21.0&  0.48   &  171$\,\pm\,5$&\\
 2016-May-21 & 16A-420$^k$ & 1604 &  B & 6.0 (C) & 2048  & 0.98, 0.86, -23.5 &   0.034&   $61\,\pm\,2$&\\
 2016-May-21 & 16A-420$^k$ & 1604 &  B & 3.0 (S) & 2048  & 1.8, 1.7, -20.0&   0.058 &   $167\,\pm\,2$&\\
 2016-May-21 & 16A-420$^k$ & 1604 &  B & 1.5 (L)& 1024 &3.6, 3.1, -20.7&  0.50   &  261$\,\pm\,5$&\\
 2019-Apr-21 &  VLASS$^k$ & 2669 &  B &  3.0 (S)&  2048 & 2.9, 2.1, 31.1 &   0.169 & 21$\,\pm\,2$&\\
 2019-Jun-24 &  19A-425$^k$ & 2733 &  B &  1.5 (L)&  64  & 4.9, 3.3, -13.8 &   0.170  & 50$\,\pm\,5$  &&\\
 \textcolor{red}{2021-Sep-14} & \textcolor{red}{21A-033$^n$} & 3546 & C\textrightarrow B & 3.0 (S)& 2048 & 2.92, 1.39, -2.83 & 0.110 & \textcolor{red}{9.53$\,\pm\,1$} &&\\
 \textcolor{red}{2022-Oct-08} & \textcolor{red}{22B-016$^n$} & 3935 & C & 3.0 (S) & 2048 & 6.55, 5.07, 42.67 & 0.04 & \textcolor{red}{19.97$\,\pm\,1$}&&\\
 \textcolor{red}{2024-Jun-25} & \textcolor{red}{VLASS$^o$} & 4561 & B & 3.0 (S) & 2048 & 3.19, 2.13, 36.00 & 0.16 & \textcolor{red}{13.8$\,\pm\,4$}&&\\
\hline
\end{tabular}

\vspace{0.2cm}

\begin{minipage}{0.95\textwidth}
\raggedright

%\begin{tablenotes}

$^a$ Date of observation.\\
$^b$ Project name or code.\\
$^c$ Elapsed time since T1 = Dec 30, 2011.\\
$^d$ VLA array configuration.\\
$^e$ Central frequency of the band followed by the name of the band in parentheses.\\
$^f$ Total bandwidth.\\
$^g$ Synthesized beam major axis, minor axis, and position angle.\\
$^h$ Measured rms map noise (primary beam corrected, near source).\\
$^i$ Flux of nucleus after correction for surrounding disk emission for low resolution images.\\
$^{j}$ Downloaded and measured from the NRAO VLA Sky Survey \citep[NVSS,][]{condon1998} \\
$^{k}$ Data from \citet{Perlman2022} \\
$^l$ Data from \citet{Irwin2015}.\\ 
$^m$ Data from \citet{perlman2017}. \\
$^n$ \textcolor{red}{This work}.\\
$^o$ \textcolor{red}{This work}. Measured from VLASS data \citep[VLASS;][]{Lacy_2020}, download form \href{https://archive-new.nrao.edu/vlass/quicklook/}{https://archive-new.nrao.edu/vlass/quicklook/}
\\
\end{minipage}
}

\label{tab:obserinf}
\end{table*}

\clearpage

\bibliography{my}{}
\bibliographystyle{aasjournalv7}

\end{document}